\documentclass[aps,epjc,twocolumn,floatfix,superscriptaddress,nofootinbib]{revtex4}
\usepackage{graphicx}
\usepackage{dcolumn}  
\usepackage{bm}       
\usepackage{palatino}
\usepackage{epsfig}
\begin{document}
\title{
Correlation function in Field - Feynman hadronization model
}
\author{L.~Grigoryan\\
Yerevan Physics Institute, Br.Alikhanian 2, 375036 Yerevan, Armenia }
\begin {abstract}
\hspace*{1em}
Very successful hadronization model for the production of mesons in the
jets of initial quarks was proposed thirty years ago by Field and Feynman.
The model is used so far in the design of experiments and for comparison
with the experimental data. In this work it is shown that correlation
function in that model, as a function of the energies of two final hadrons
has interesting properties (energies are taken in fractions of initial
quark energy, and equal $z_{1}$ and $z_{2}$, respectively). In particular,
for some pairs of mesons there are kinematic regions over variable $z_{2}$
(at fixed values of $z_{1}$), where it is positive and unexpectedly large.
The experimental investigation of the correlation functions in these
kinematic regions has twofold interest, from one side it can help to
investigate some delicate properties of quark jets and from the other side
can help to verify the physical suppositions of model, and to obtain more
precise values of parameters.
\end {abstract}
\maketitle
\section{Introduction}
\normalsize
\hspace*{1em}
In the paper~\cite{field1} (hereafter called FF1), the authors investigated
the consequences of the assumption that the high-transverse-momentum
particles seen in hadron-hadron collisions are produced by a single, hard,
large-angle elastic scattering of quarks, one from the target and one from
the beam. The fast outgoing quarks were assumed to fragment into a cascade
jet of hadrons. The distribution of quarks in the incoming hadrons were
determined from lepton-hadron inelastic scattering data, together with
certain theoretical constraints such as sum rules, etc. The manner in which
quarks cascade into hadrons was determined from particle distributions seen
in lepton-hadron and lepton-lepton collisions supplemented by theoretical 
arguments. Parameterizations for Single-hadron Fragmentation Functions (SFF)
were obtained. In the next work~\cite{field2} a new, much simpler,
parameterization for these functions was provided.
The answers to many other questions concerning the details
of the hadronization process were given.
This hadronization model for the production of mesons in the jets of initial
quarks and antiquarks turned out very successful and is used so far in the
design of experiments in which quark jets may be observed and for comparison
with the available experimental data. It provides a parameterization of the
properties of the jet of mesons generated by a fast outgoing quark. It is
assumed that the meson that contains the original quark leaves momentum and
flavor to a remaining jet in which the particles are distributed (except for
scaling of the energy and possible changes of flavor) like those of the
original jet. One function, the probability $f(\eta)$ that the remaining jet
has a fraction $\eta$ of the momentum of the original jet, is chosen (as a
parabola) so the final distribution of charged hadrons agrees with data from
lepton experiments. All the properties of quark jets are determined from
$f(\eta)$ and three parameters: a) the degree that SU(3) is broken in the
formation of new quark-antiquark pairs ($s\bar{s}$ is taken as half as likely
as $u\bar{u}$) $u\bar{u}:d\bar{d}:s\bar{s} = \gamma_{u}:\gamma_{d}:\gamma_{s}
= 1:1:\frac{1}{2}$ (as usually, it is adopted the well-known condition
following from the isospin symmetry $\gamma_{u} = \gamma_{d}$) and for the
sum of probabilities it is taken $\gamma_{u} + \gamma_{d} + \gamma_{s} = 1$,
because the probabilities of the production of the heavier quarks are neglected;
b) the spin nature of the primary mesons (in the basic version of model is
assumed to be vector and
pseudoscalar with equal probability) $\alpha_{v} = \alpha_{ps} = \frac
{1}{2}$; 
c) the mean transverse momentum given to these primary mesons\footnote{In this
work we do not consider the transverse momenta of hadrons and suppose that
integration over them are performed.}.
In the rest of the paper we will frequently refer to the Ref.~\cite{field2}
and for this reason it is convenient to denote its as FF2, and model, which
was developed there to denote as Field-Feynman Quark Jet Model (FFQJM).\\
\hspace*{1em}
We would like present here the explanation of the success of the FFQJM,
which was given by Gottschalk in his lectures~\cite{gottschalk}:
"Specifically, the FFQJM is not a "theory" of how hadronization works.
Rather, it is simply a parameterization of hadron formation in jets which
reproduces two important experimental facts: 1) approximate scaling of the
energy distributions of hadrons within a jet; 2) limited transverse momenta of
the produced hadrons with respect to the jet axis. These qualitative features
of hard scattering data are not "explained" or "derived" in the FFQJM. Rather,
the FFQJM is a parameterization of hadron production from fast quarks which is
constructed in such a manner that the above experimental observations are
reproduced."\\
\hspace*{1em}
Let us briefly discuss the parameters from points a) and b) of FFQJM presented
above. They
play important role in the understanding of the hadronization process. In the
early model of Field and Feynman presented in FF1, the point a) coincides with
that from FFQJM, but point b) differs. For its in FF1 is chosen simple version
$\alpha_{ps} = 1$, $\alpha_{v} = 0$.
In the simple Lund model~\cite{simpleL}, which has very large similarities with
the FFQJM, the point a) coincides with latter,
but for parameters of point b) the ratio is chosen in the form
$\alpha_{v}:\alpha_{ps} = 3:1$ as given by statistical spin counting.
In the later version of Lund model, the so called standard Lund model
~\cite{standardL} the point a) is changed to $\gamma_{u}:\gamma_{d}:\gamma_{s}
= 1:1:\frac{1}{3}$ which is followed from the tunneling production
mechanism, but point b) coincides with one from FFQJM,
because authors state that the spin-spin forces enhance the production
of the lighter pseudoscalars, from the naive $3:1$ vector to pseudoscalar
ratio to something closer to $1:1$.  
Later these parameters were investigated in detail
in the process of $e^{+}e^{-}$ annihilation (see, for instance,
Refs.~\cite{anselmino} and~\cite{chliapnikov}). In~\cite{chliapnikov} it is shown
that the production rates of light-flavor hadrons are set by their masses,
spins, strangeness suppression, and strongly influenced by the spin-spin 
interactions of their quarks. The vector-to-pseudoscalar meson and 
decuplet-to-octet baryon suppresions are the same and are explained by the
hyperfine mass splitting. The strangeness suppression factor,
$\lambda = 2\gamma_{s}/(\gamma_{u} +\gamma_{d}) = 0.295 \pm 0.006$,
is the same for mesons and baryons. It is related
to the difference in the constituent quark masses,
$\lambda = e^{-(m_{s}-m_{l})/T}$, where $m_{l} = m_{u} = m_{d}$, at the
temperature, $T = 142.4 \pm 1.8 MeV$. 
Comparison of the parameters from FFQJM with the ones
obtained in other models showes that their values
differ significantly. This means, that values of parameters in FFQJM were
determined with large uncertainties.
Inclusion in consideration the new sets of experimental data
and new observables can improve description and shift the parameters to
their precise values.\\
\hspace*{1em}
In FFQJM, using formalism based on the recursive principle and available
experimental data, were obtained SFF $D_{q}^{h}(z)$, which are the
distributions of hadrons of flavor
\footnote{Following FF2 we will call the isospin and strangeness
properties the "flavor" of the primary mesons.}
$h$ from a quark $q$, which carry away the fraction $z$ of its energy,
and Dihadron Fragmentation Functions (DFF)
$D_{q}^{h_{1}h_{2}}(z_{1},z_{2})$,
which are the distributions of hadrons of flavors
$h_{1}$ and $h_{2}$ from a quark $q$,
which carry away the fractions $z_{1}$ and $z_{2}$ of its energy,
respectively. Using these results, authors proposed several experiments,
which can verify model. Among others, they proposed to measure the
correlation functions\footnote{It is worth to mention, that the correlations
between final hadrons in the framework of bootstrap model, based on the
recursive principle, was discussed, for the first time, in Ref.~\cite{jengo}.
We are grateful to Professor A.Krzywicki for the comment concerning this
point.}
for oppositely and identically charged hadrons.
It was predicted, that correlation function for the oppositely charged
hadrons must show the characteristic short-range correlation behavior.
The correlation between two identically charged hadrons, on the
other hand,was predicted to be quite small. In FF2 the correlation function
of two hadrons was presented as a function of their "$z$ rapidity" given
by $Y_{z} = -ln(z)$. Were considered energies large enough that a plateau
was developed. Rapidity of the first hadron was fixed in region of plateau,
$Y_{z_{1}} = 4.0$ (which corresponds $z_{1} \approx 0.02$) and $Y_{z_{2}}$ was
changed in wide region $0 < Y_{z_{2}} <6$. As it is well known,
in this model energy and momentum of the quark jets do not precisely
conserved. The violation of the energy conservation reaches a few per cents.
In these conditions the consideration of hadrons with small enough energy
may be doubtful.\\
\hspace*{1em}
In this work we study the correlation function in the framework of
FFQJM in detail, for the one definite reaction, the
semi-inclusive electroproduction
of hadrons on proton in Deep-Inelastic Scattering (DIS) region, at
moderate energies (energy of the virtual photon or, which is the same,
the energy of the initial quark in order of 10-20 GeV),
where the plateau does not formed and where the small
values of $z$ in order of $0.02$ do not belong to the current
fragmentation region. Instead of pairs of oppositely and
identically charged hadrons we consider pairs of hadrons having definite
charge and flavor; and instead of rapidity, which has very narrow changing
region at moderate energies, we use variable $z$.
It is shown that the correlation function has interesting
properties. In particular for some pairs of hadrons there are kinematic
regions over variable $z_{2}$ (at fixed values of $z_{1}$)
where it is positive and unexpectedly large.
The experimental investigation of the correlation
function in these kinematic regions has twofold interest, from one side
it can help to investigate some delicate properties of quark jets and from
the other side can help to verify the physical suppositions
of model, and to obtain more precise values of parameters.\\
\hspace*{1em}
There are, of course, several obvious defects of the model, which were
discussed by various people begun with authors of model,
Field and Feynman. The detailed
criticism of model can be found in work~\cite{gottschalk}.
There are, also, many attempts to improve the
model for the more precise description of the experimental data
(see, for instance, Refs.~\cite{gottschalk} and~\cite{jing_hua}).\\
\hspace*{1em}
However, despite on the available defects, the model is able to describe
the experimental data in DIS region with good enough precision, as
the numerous applications are shown. Therefore, we think that it
can be used for the qualitative study of the two
hadron correlation functions at moderate energies.\\
\hspace*{1em}
And last point, which we would like to discuss here, is the
place of this work among others, devoted to the investigation of the
DFF. At present, many questions connected
with the hadron-hadron correlation become the objects of the experimental
investigations in the high-energy lepton-lepton, lepton-hadron,
lepton-nucleus, hadron-hadron, hadron-nucleus and nucleus-nucleus
interactions. As a consequence are arisen theoretical works, which devoted
to the explanation and description of the existing experimental data or
to the proposals for future experiments. There is huge amount
of such experimental and theoretical works. Naturally it is impossible
to present them in one article. We present only two of them, because they
reflect common approach to the investigation of DFF on phenomenological
level. They are Refs.~\cite{majumder1}, and~\cite{majumder2}. In these
works authors had concentrated on the definition of the DFF and derived the
DGLAP evolution equations for the non-singlet quark DFF~\cite{majumder1}
and singlet quark and gluon DFF~\cite{majumder2}. The more
attention was paid to the derivation of DGLAP evolution equations and less
to the derivation of proper initial conditions. About initial conditions used
in~\cite{majumder1}, authors themselves note, that in numerical study of the
nonsinglet quark DFF they used a simple ansatz for the initial condition
as $D_{NS}(z_{1},z_{2}) = D(z_{1}) \times D(z_{2})$, which at best is just
a guess and differs significantly from the inherent hadron correlations in
a single jet. In the next work~\cite{majumder2} the initial conditions
for the DFF were extracted from JETSET at a scale $Q_{0}^{2} = 2GeV^{2}$.
Although both the SFF and DFF evolve rapidly with $Q^{2}$, their ratio,
which frequently is used for the comparison with experimental data,
has a very weak $Q^{2}$ dependence. This is especially true for the ratio
in the case of the gluon fragmentation function, which shows practically
no change with $Q^{2}$. The results of evolution are strongly dependent,
however, on the initial conditions and thus on the actual values of $z_{1}$
and $z_{2}$. In this work we study the DIS at moderate
energies and $Q^{2}$, i.e. we investigate initial conditions for DFF for
different pairs of hadrons and different values of $z_{1}$ and $z_{2}$.
We show, that different choice of the types of hadrons and their fractional 
energies leads to the essentially different initial conditions for DFF.\\ 
\hspace*{1em}
The paper is organized as follows. In Section 2 we briefly discuss 
the structure of FFQJM and point out the way of the derivation of the
formulae for the calculations of SFF and DFF, and present expression for
correlation function.
Section 3 presents Results and Discussion. Our Conclusions are
presented in Section 4.
\section{Theoretical framework}
\normalsize
\hspace*{1em}
Let us consider the semi-inclusive DIS process on proton, in which a
two hadron system is observed in the final state:
\begin{eqnarray}
e + p \to e^{'} + h_{1} + h_{2} + X \hspace{0.3cm}.
\end{eqnarray}
We would like to study correlation between two hadrons in the framework of the
FFQJM.
Before we want briefly remind, that main ingredients of FFQJM are 
the probability function $f(\eta) = 1 - a + 3a\eta^{2}$ and three free parameters:
1) the ratio $\gamma_{s}/\gamma_{u}$; 2) the probability of pseudoscalar
meson production $\alpha_{ps} (\alpha_{ps} + \alpha_{v} = 1)$
and 3) the mean transverse momentum of the primary mesons.
As we already mentioned in Introduction, we do not consider third parameter
and suppose that integration over it is performed. It is easily to see,
that parameters from points 1 and 2 are correlated with parameter $a$ entering
in function $f(\eta)$. In FFQJM the ratio $\gamma_{s}/\gamma_{u}$ is fixed on value
$0.5$, the choice of parameter $\alpha_{ps}$ and correlated with it parameter
$a$ leads to the two sets of parameters (two versions of model) considered in
FF2. They are: a) simple version, which takes into account the direct production
of pseudoscalar mesons only, with corresponding set of parameters $a = 0.88,
\alpha_{ps} = 1, \alpha_{v} = 0$; and b) basic version, which takes into account 
both the direct production of pseudoscalar mesons and their production from the 
decays of resonances, with set of parameters $a = 0.77,
\alpha_{ps} = \alpha_{v} = 0.5$.\\
\hspace*{1em}
Now we introduce the notion "rank" following FF2. Let us consider the "hierarchy"
structure of the final mesons produced when a quark of type "a" fragments into
hadrons. New quark pairs $b\bar{b}$, $c\bar{c}$, $d\bar{d}$ etc., are produced and
"primary" mesons are formed. The "primary" meson $a\bar{b}$ that contains the
original quark is said to have "rank" one, the "primary" meson $b\bar{c}$ "rank"
two, the "primary" meson $c\bar{d}$ "rank" three, etc. Finally, some of the
"primary" mesons decay and it is assigned, that all the decay products to have the
"rank" of the parent. The order in "hierarchy" is not the same as order in momentum
or rapidity.\\
\hspace*{1em}
As it was mentioned above we use for SFF and DFF formulae obtained in FF2.
Unfortunately, corresponding expressions are very long and inconvenient for
reading. We do not presented these formulae here, and only remind the numbers
of corresponding equations in FF2. For SFF it is used formula (2.57)
from FF2. The structure of DFF is more complicate and consists of five
terms: (i) the probability that the primary meson at $z_{1}$ is of rank 1
and the primary meson at $z_{2}$ is of rank 2, given by (2.43a);
(ii) the probability that $z_{1}$ is of rank one, but $z_{2}$ is of rank
higher than 2, given by (2.43b); (iii) the probability that the primary meson
at $z_{1}$ is not first in rank but higher, but the primary meson at $z_{2}$
is directly of next rank to the one at $z_{1}$, given by (2.43c);
(iv) neither the primary meson at $z_{1}$ or $z_{2}$ is first in rank, nor
are they adjacent, given by (2.43d). As it is pointed out in FF2, the
complete DFF for producing two hadrons of flavor $h_{1} = a\bar{b}$ and
$h_{2} = c\bar{d}$ is given by symmetrizing (2.43a-d) with respect to
$z_{1}$ and $z_{2}$ (see eq. (2.44) of FF2). As pointed out in FF2,
inclusion in the consideration the mesons produced from the decays of
resonances complicates the structure of equations (2.43a-d) and
adds next term (v) corresponding to the contribution from the case where
both the mesons at $z_{1}$ and $z_{2}$ came from the decay of the same
resonance (type $h_{v}$) given by formula (2.59) of FF2.\\
\hspace*{1em}
There are two sources of correlations in the discussed model. Naturally,
there is the correlation among secondary particles that are the decay
products of the same primary meson. In addition, however, the primary
mesons are not formed at random in $z$. Primary mesons adjacent in
rank are correlated in both flavor and $z$ since they each contain
a quark (or antiquark) that came from the same $q\bar{q}$ pair. The two
primary mesons of adjacent rank tend to occur near each other in $z$.
All flavor correlations in the quark jets occur between primary mesons
of adjacent rank. The flavor of a meson of rank $r + 2$ is independent
of the flavor of the meson of rank $r$.\\
\hspace*{1em}
The observable which is widely used for the investigation of correlation
between two hadrons produced in the any hard process is
the two-body correlation function. It is a function of single and double hadron
multiplicities $\rho^{h}_{1}(z)$ and $\rho^{h_{1}h_{2}}_{2}(z_{1}, z_{2})$,
respectively. In our case $h_{i}$ denote the hadrons, $i = 1, 2$; $z_{i}$
denote the fraction of the virtual photon energy $\nu$ carried by $i$-th
hadron, $z_{i} = E_{i}/\nu$ where $E_{i}$ is the energy of the $i$-th hadron.
In this work we will use, as observable, the "normalized" two body correlation
function (see, for instance, FF2 and references therein)
\begin{eqnarray}
R_{cor} = R^{h_{1}h_{2}}(z_{1}, z_{2}) =
\frac{\rho^{h_{1}h_{2}}_{2}(z_{1}, z_{2})}{\rho^{h_{1}}_{1}(z_{1})
\rho^{h_{2}}_{1}(z_{2})} - 1 \hspace{0.3cm}.
\end{eqnarray}
In the quark-parton model the single hadron multiplicity is
\begin{eqnarray}
\rho^{h}_{1}(z) = \frac{1}{\sigma}\frac{d\sigma^{h}}{dz} =
\frac{\sum_{i}e^{2}_{i}q_{i}(x_{Bj})D^{h}_{i}(z)}{\sum_{i}e^{2}_{i}q_{i}(x_{Bj})}
\hspace{0.3cm},
\end{eqnarray}
where $\sigma$ and $\frac{d\sigma^{h}}{dz}$ are inclusive and semi-inclusive
cross sections, $e_{i}$ and $q_{i}(x_{Bj})$ are the electric charge of the
quark in units of the elementary charge and distribution function
of the quark with flavor $i (i = u, d, s)$ in proton,
$x_{Bj}$ is the Bjorken variable $x_{Bj} = Q^{2}/2M_{p}\nu$, where
$Q^{2}$ is the photon virtuality and $M_{p}$ is the proton mass.
$D^{h}_{i}(z)$ is the SFF for the
production of hadron $h$ by $i$-th quark, $z$ denote the fraction of the
virtual photon energy $\nu$ carried by hadron.
\begin{eqnarray}
\rho^{h_{1}h_{2}}_{2}(z_{1}, z_{2}) = \frac{1}{\sigma}\frac{d^{2}\sigma^{h_{1}h_{2}}}
{dz_{1}dz_{2}} =
\frac{\sum_{i}e^{2}_{i}q_{i}(x_{Bj})D^{h_{1}h_{2}}_{i}(z_{1},
z_{2})}{\sum_{i}e^{2}_{i}q_{i}(x_{Bj})},
\end{eqnarray}
where $\frac{d^{2}\sigma^{h_{1}h_{2}}}{dz_{1}dz_{2}}$ is the
semi-inclusive cross section in which two hadrons in the final state are
observed. The DFF
$D_{i}^{h_{1}h_{2}}(z_{1},z_{2})$, are the distributions of hadrons of flavors
$h_{1}$ and $h_{2}$ from a quark $i$, which carry away the fractions $z_{1}$
and $z_{2}$ of its energy, respectively.
\section{Results and Discussion}
\normalsize
\hspace*{1em}
We perform calculations mainly for basic version of the model. In
this case it is necessary to take into account that a part of the final
pseudoscalar mesons are arisen in result of decays of the other hadrons.
We use contributions from 13 hadrons ($K^{0}$, $\bar{K^{0}}$, $\rho^{+}$,
$\rho^{0}$, $\rho^{-}$, $\eta$, $\eta^{'}$, $K^{*+}$, $K^{*0}$, $K^{*-}$,   
$\bar{K}^{*0}$, $\omega$, $\phi$). Further we will call they "resonances"
although among them are long living hadrons also.
Some of them, in result of decay,
give contribution only in pion or kaon production, but there are resonances
which decay into $\pi K$ pairs.
From eqs.(2)-(4) we are obtained the correlation functions $R_{cor}$ for
the pairs of mesons with different types and/or electric charges
($\pi \pi$, $\pi K$ and $K K$). For calculations were used expressions for
SFF and DFF obtained in the framework of the
FFQJM and the distribution functions of the quarks in the proton
from~\cite{glueck}. For self-consistency of the consideration we use
distribution functions of the quarks in the proton in the Leading Order (LO)
approximation. For comparison with the basic version, we will present
results of calculations for simple version of model also.\\
\begin{figure}[!ht]
\begin{center}
\epsfxsize=8.cm
\epsfbox{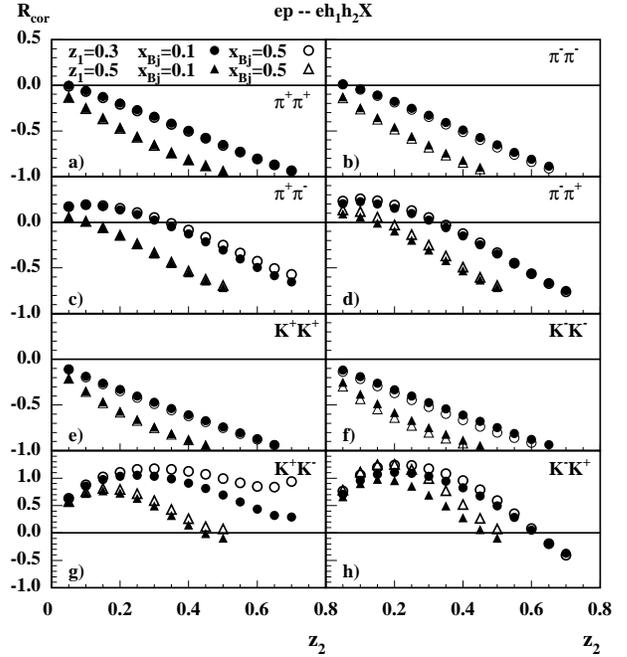}
\end{center}
\caption{\label{xx1}
{\it
Correlation functions $R_{cor}$ for pairs of hadrons of the same type
($\pi \pi$ and $K K$), with identical and opposite electric charges
as a function of $z_2$ are presented. Calculations were performed for
a fixed values of $z_1$ equal to $0.3$ (circles) and $0.5$ (triangles).
At each fixed value of $z_1$ two curves with fixed values of $x_{Bj}$
equal to $0.1$ (filled symbols) and $0.5$ (open symbols) are presented.
The correlation functions were calculated in basic version of the FFQJM,
taking into account hadrons produced both directly and from resonance
decays. Only pairs of hadrons with different electric charges have positive  
correlation function.
}}
\end{figure}
\begin{figure}[!ht]
\begin{center}
\epsfxsize=8.cm
\epsfbox{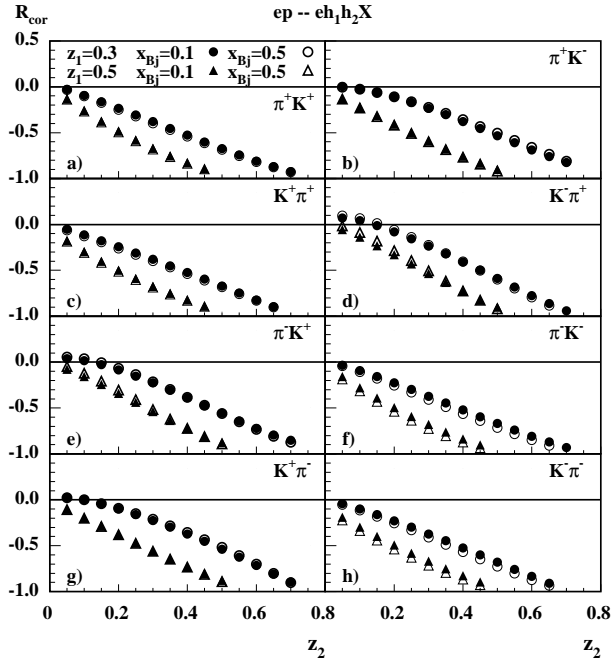}
\end{center}
\caption{\label{xx2}
{\it
The same as in Fig.1 for the pairs of hadrons of different types, having      
identical and opposite electric charges. Correlation functions are negative
practically everywhere.
}}
\end{figure}
\begin{figure}[!ht]
\begin{center}
\epsfxsize=8.cm
\epsfbox{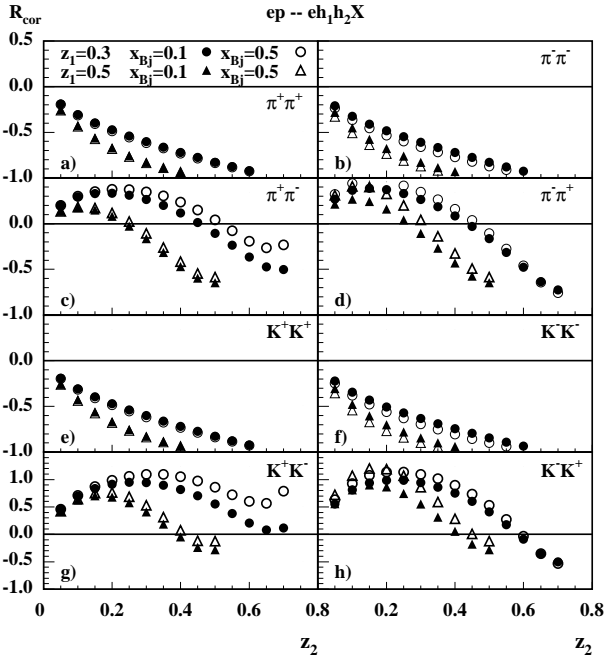}
\end{center}
\caption{\label{xx3}
{\it
The same as in Fig.1 for correlation functions calculated in the
simple version of the FFQJM, taking into account only
hadrons produced directly. Easily to see, that in comparison with basic
version, absolute values of $R_{cor}$ are here bigger.
}}
\end{figure}
\hspace*{1em}
In Fig.1 the
correlation functions $R_{cor}$ for the pairs of hadrons of the same type
($\pi \pi$ and $K K$), with identical and opposite electric charges
as a functions of $z_2$ are presented. Calculations were performed for  
a fixed values of $z_1$ equal to $0.3$ (circles) and $0.5$ (triangles).
At each fixed value of $z_1$ two curves with fixed values
of $x_{Bj}$\footnote{We can not take for $x_{Bj}$ too small values, because
then become large the contributions of two jet events, which connected with
the Pomeron exchange. In other words, it is need take into account that
the part of its time in order of $1/x_{Bj}M_{p}$ the virtual photon behaves
as a $q\bar{q}$ pair. If this time is essentially larger than the size of
proton then instead of the photon with target will interact quark-antiquark
pair. We can escape this situation imposing the restriction $x_{Bj} \ge 0.1$.}
equal to $0.1$ (filled symbols) and $0.5$ (open symbols) are presented.
The correlation
functions are calculated in the basic version of the FFQJM, taking into
account hadrons produced both directly and from decays of resonances. Only
for the pairs of hadrons of the same type and different electric charges
they have positive values (see panels c, d, g, h).
The correlation functions for $\pi^{+} \pi^{-}$ and $\pi^{-} \pi^{+}$ pairs
have similar shapes and become positive only at smaller value of
$z_{1}$ using in calculations ($z_{1} = 0.3$).
At bigger value of $z_{1}$ using in calculations ($z_{1} = 0.5$)
they are negative. In the cases of $K^{+} K^{-}$ and $K^{-} K^{+}$ pairs   
we see more strong positive correlation (pay attention on the scale of
corresponding panels). Positive correlation persists practically over  
all range of the change of variables. Next peculiarity of the correlation
functions for these pairs
is the strong enough $x_{Bj}$ - dependence. In case of $K^{+} K^{-}$ more
strong $x_{Bj}$ - dependence takes place for $z_{1} = 0.3$, while in case of
$K^{-} K^{+}$ it does for $z_{1} = 0.5$. Pairs of hadrons with identical
flavors (having the same type and identical electric charges)
($\pi^{+} \pi^{+}$, $\pi^{-} \pi^{-}$, $K^{+} K^{+}$ and $K^{-} K^{-}$)
have correlation functions which are negative in all region of calculated
$z_{2}$. They decrease (practically linearly) in all region of $z_{2}$.
Later we will discuss the reasons of the different behavior of the pairs of
hadrons of the same type, for cases, when they have opposite and 
identical electric charges.\\
\hspace*{1em}
In Fig.2 the correlation functions for the pairs of hadrons with different
types ($\pi K$ - pairs), with identical and opposite electric charges are
presented. Notations are the same as in Fig.1. The correlation functions are
calculated in basic version of the FFQJM. They are negative practically
everywhere. Their behavior over variable $z_{2}$ are close to the linear. We
see only small differences in the shapes and values. The $x_{Bj}$ - dependence
is practically absent.\\
\hspace*{1em}
In Fig.3 the   
correlation functions $R_{cor}$ for the pairs of hadrons of the same type
($\pi \pi$ and $K K$), with identical and opposite electric charges
as a function of $z_2$ are presented. Notations are the same as in Fig.1.
The correlation functions are calculated in the   
simple version of the FFQJM, taking into account only
hadrons produced directly. Easily to see, that in comparison with basic
version, presented in Fig.1, absolute values of $R_{cor}$ here bigger.
The qualitative
behavior of the correlation functions in basic and simple versions of model
is very close. The positive correlation for $\pi^{+} \pi^{-}$ and
$\pi^{-} \pi^{+}$ is more prominent in simple version, while for $K^{+} K^{-}$
and $K^{-} K^{+}$ pairs the situation is practically the same in both versions.\\
\begin{figure}[!ht]
\begin{center}
\epsfxsize=8.cm
\epsfbox{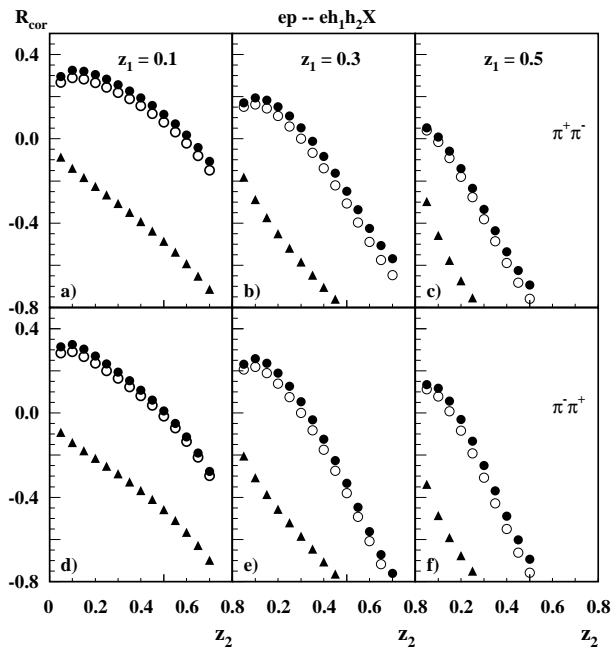}
\end{center}
\caption{\label{xx4}
{\it
Correlation functions $R_{cor}$ for $\pi \pi$ pairs with opposite
electric charges as a functions of $z_2$ are presented.
Calculations were performed for
a fixed values of $z_1$ equal to $0.1$, $0.3$ and $0.5$.
The correlation functions were calculated in basic version of the
FFQJM, taking into account hadrons produced both directly
and from resonance decays.
It is presented the contributions of the different mechanisms in
correlation functions. The contribution of not adjacent hadrons only
(filled triangles); the contribution of not adjacent and adjacent
hadrons (open circles); the contribution of not adjacent and adjacent
hadrons and hadrons produced from the same resonance (filled circles).
}}
\end{figure}
\begin{figure}[!ht]
\begin{center}
\epsfxsize=8.cm
\epsfbox{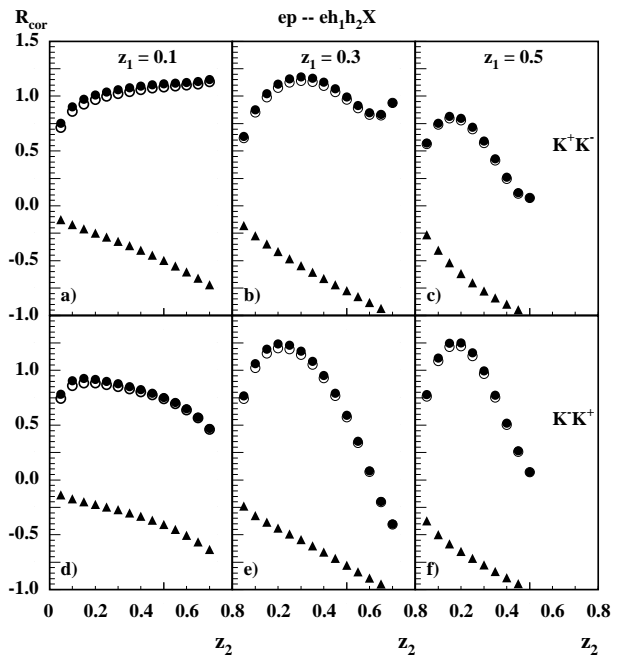}
\end{center}
\caption{\label{xx5}
{\it
The same as in Fig.4 for $K K$ pairs. (Attention: other scale!)
}}
\end{figure}
\hspace*{1em}
In Fig4. the correlation function $R_{cor}$ for $\pi \pi$ pairs with opposite
electric charges as a function of $z_2$ is presented. Calculations were performed
for a fixed values of $z_1$ equal to $0.1$, $0.3$ and $0.5$. The correlation
function is calculated in the basic version of the FFQJM, taking into account
hadrons produced both directly and from decays of resonances. It is presented the 
contributions of the different mechanisms in correlation function. The
contribution of not adjacent in rank hadrons only (filled triangles); the sum of 
contributions of not adjacent and adjacent hadrons (open circles); and final
result, the sum of three contributions: of not adjacent hadrons, adjacent
hadrons and hadrons produced from the same resonance (filled circles).\\
\hspace*{1em}
In the Fig.5 the same as in Fig.4 for $K K$ pairs is presented (pay attention
on the scale).\\
\hspace*{1em}
Now let us discuss briefly the Figs.4 and 5. First observation is, that taking
into account only contributions from not adjacent mesons we obtain for pairs
$\pi^{+} \pi^{-}$, $\pi^{-} \pi^{+}$, $K^{+} K^{-}$ and $K^{-} K^{+}$ the negative
correlation functions which have the behavior very close to the cases for pairs
of mesons of the different types, or for pairs of mesons of the same type and
electric charges (identical flavors).
The main source of the positive correlation is the possibility for the mesons
entering in the pairs $\pi^{+} \pi^{-}$, $\pi^{-} \pi^{+}$, $K^{+} K^{-}$ and
$K^{-} K^{+}$ to be produced in the adjacent rank. Second source of the positive
correlation, the possibility for the mesons be produced from the decay of the same
resonance is very small and can be neglected. From panels a, b, c of Fig.4 for
$\pi^{+} \pi^{-}$ pair, we see that while the contribution of not adjacent mesons
is decreased with the increasing of $z_{2}$, the contribution of adjacent mesons
is practically constant in all region of $z_{2}$. The similar behavior takes place
for $\pi^{-} \pi^{+}$ pair (see panels d, e, f of Fig.4). Situation is more 
complicated for the $K^{+} K^{-}$ and $K^{-} K^{+}$ pairs (see Fig.5). The
behavior of the contributions from not adjacent mesons is like the $\pi \pi$ case 
in Fig.4. The contribution of the adjacent mesons for $K K$ pairs has more 
complicate behavior as a function of $z_{1}$ and $z_{2}$ and at some values of 
$z_{1}$ and $z_{2}$ can be essentially larger than the contribution from not 
adjacent mesons. It is the source of the big and nontrivial positive correlations 
for the $K^{+} K^{-}$ and $K^{-} K^{+}$ pairs.\\
\hspace*{1em}
It is very important to verify, that results obtained above do not depend
essentially from the choice of the parton distribution functions. For this
goal we repeated all calculations with another set of parton distribution
functions. Were used LO parton distribution functions from~\cite{gehrmann}. It
was obtained, that difference between two results is smaller than 1 per cent.\\ 
\hspace*{1em}
We propose to measure the correlation functions for the $K^{+} K^{-}$ and
$K^{-} K^{+}$ pairs experimentally. The experimental discovery of the fact of
big positive correlation will means, that main mechanism of hadronization is
the fragmentation of the initial quark via consecutive production of $q\bar{q}$
pairs and their further combination into final mesons. 
\section{Conclusions}
\normalsize
\hspace*{1em}
In the absence of any dynamical
correlations, the probability to observe in a single jet one hadron
with fraction of energy $z_{1}$ and second hadron with $z_{2}$, together
with anything else, would be equal to the product of the probabilities
to find hadrons with $z_{1}$ and $z_{2}$ in different jets and $R_{cor}$,
as we defined it in eqs.(2)-(4), would be close to zero. Really, it is true
only in case $z_{1} + z_{2} \ll 1$. When $z_{1} + z_{2} \sim 1$ the
correlation function for the pair of hadrons, which can not be produced
adjacent in rank becomes negative, because of the necessity of the production
of the additional particles. Even in the case, when hadrons can be produced 
adjacent in rank part of energy is lost because the additional hadrons are
produced in result of decays of resonances, which also leads to the reduction
of correlations. We obtained such a behavior for all considered pairs of mesons,
besides $\pi^{+} \pi^{-}$, $\pi^{-} \pi^{+}$, $K^{+} K^{-}$ and $K^{-} K^{+}$.
The prominent positive correlations for the $K^{+} K^{-}$ and $K^{-} K^{+}$
pairs in the framework of FFQJM were obtained. The experimental investigation
of $R_{cor}$ for these cases can help to verify the basic assumptions of the
model and receive the valuable information about hadronization process.\\
\hspace*{1em}
The model does not include quantum effects. It has dealt solely with
probabilities and not amplitudes. It is supposed that in DIS region the
interference effects are small. If the interference effects, nevertheless,
do not small, it will felt in selected pairs $\pi^{+} \pi^{-}$, $\pi^{-} \pi^{+}$
because of the interference between the amplitudes for the production of the 
hadrons $h_{1}$ and $h_{2}$ adjacent and not adjacent in rank. For the pairs
$K^{+} K^{-}$ and $K^{-} K^{+}$ the interference will be practically impossible,
because of the necessity of the production of the additional pair of $s\bar{s}$,
and construction of specific final state.\\
\hspace*{1em}
The advatage of our approach to the derivation of SFF and DFF
in comparison with, for instance,~\cite{majumder1},
and~\cite{majumder2} is that we have, in the framework of FFQJM,
the analytic formulae for the SFF and DFF, which are obtained at moderate values
of $Q^{2}$. They can serve as a initial state SFF and DFF for the QCD evolution.
Moreover we can have separately the contributions in DFF
from different mechanisms (production of hadrons adjacent and not adjacent in rank,
production from the decay of the same resonance, etc.). It is very interesting
to perform the QCD evolution for the different initial states and to trace the 
influence of the choice of the initial state, on the final state at 
high $Q^{2}$.\\
\hspace*{1em}
As we mentioned above,
the main source of the positive correlation is the possibility for the mesons
entering in the pairs $\pi^{+} \pi^{-}$, $\pi^{-} \pi^{+}$, $K^{+} K^{-}$ and
$K^{-} K^{+}$ to be produced in the adjacent rank. This mechanism is not the
peculiarity of the FFQJM and is available in all hadronization models of the
fragmentation type (for instance~\cite{simpleL} and~\cite{standardL}). It will
be very interesting item to receive the correlation functions in other
hadronization models.\\
\begin {acknowledgments}
\hspace*{1em}
The author wish to thank N. Akopov, Z. Akopov and H. Gulkanyan
for helpful discussions.
\end {acknowledgments}

\end{document}